\documentclass[manuscript]{aastex}

\received{}
\revised{}
\accepted{}

\shortauthors{Jones \& Amini}
\shorttitle{Magnetic Field in DR21}

\begin{document}

\title{The Magnetic Field Geometry in DR21}
\author{Terry Jay Jones\altaffilmark{1} and Hassib Amini}

\affil{Department of Astronomy, University of Minnesota}
\affil{Minneapolis, MN 55455} 
\email{tjj@astro.spa.umn.edu}

\altaffiltext{1}{Visiting Astronomer at the Infrared Telescope
Facility which is operated by the University of Hawaii under contract from the National Aeronautics and Space Administration. }

\begin{abstract}
We present broadband imaging polarimetry of DR21 at 2.2$\mu$m. Background stars shining through the lobes of the bipolar outflow show polarization aligned with the long axis of the outflow, indicating a magnetic field geometry oriented along the flow axis. There is no indication of a spiral or turbulent magnetic field geometry in the lobes. The polarization of stars in the central cluster has a different position angle than the lobes and is in good agreement with millimeter polarimetry. The nebulosity in the Eastern lobe has moderate to high polarization consistent with scattering of continuum light from the central cluster. We were unable to detect polarization of the nebulosity in the Western lobe at the 4.2\% (3$\sigma$) level.
\end{abstract}


\section{Introduction}

The driving mechanism for the bipolar outflows so often seen associated with Young Stellar Objects (YSOs) is still a subject of considerable debate. \citet{kon99} reviewed efforts to extend the commonly invoked centrifugally driven wind models of bipolar outflows in low-mass stars to high mass YSO's. In these models the magnetic field is twisted into a spiral pattern rising vertically off of the star's dense, rotating circumstellar disk. Material entrained in this field is forced up the vertical axis and produces the observed collimated outflow \citep{pud98, kud97}. 

Determining the magnetic field geometry in bipolar YSO outflows is a difficult observational task. These outflows are most easily studied when the outflow axis lies in the plane of the sky and the star's equatorial disk is viewed nearly edge-on. This projection produces the classic bow-tie morphology seen in so many sources and allows both poles of the outflow to be easily separated spatially on the sky. If dust grains in the bipolar lobes are aligned by the magnetic field in the outflow, then polarimetry of starlight shining through the outflow can measure the projected magnetic field geometry. If most of the material in the flow contains a spiral magnetic field with a high pitch angle, then we may be able to observationally distinguish the magnetically driven models from a model that invokes some other flow mechanism in which the field is simply dragged along radially outward. 

Since these outflows can be dusty and usually occur in star forming regions, infrared polarimetry of background starlight shining through the outflow can more easily reveal the composite magnetic field geometry along that line of sight than optical polarimetry. Traditional aperture photopolarimetry of field stars can be compromised by contamination from scattered light (reflection nebulosity) in the beam. By using imaging polarimetry, a diffuse component in the images can be effectively removed from 'under' the star's image. Imaging polarimetry also has the advantage that an image will usually contain several stars that can be measured. Traditional aperture polarimetry in the red (e.g. R and I bands) has been done for several sources and generally shows alignment of the local magnetic field with the outflow direction \citep[cf.]{hod90}.

A related technique has been employed by \citet[hereafter ICBHT]{ito99}. They used narrow band imaging polarimetry of knots of molecular hydrogen emission in the massive outflow associated with DR21 to probe the projected magnetic field geometry there. DR21 is one of the most massive and powerful bipolar outflows from young stars known. As such, it is an ideal laboratory for investigating the origin of the outflow. Their technique has the advantage of using emission that is from within the outflow itself, uncontaminated by transmission through background dust. However, it may suffer from contamination by underlying reflection nebulosity that could be hard to remove.

In this paper we report imaging polarimetry of DR21 in the broadband K (2.2$\mu$m) filter. We observe both background field stars shining through the outflow and reflection nebulosity in the flows. Note that background starlight samples all of the dust along the line of sight through the outflows, so we can not distinguish between material entrained by the outflow in a surrounding sheath and material interior to the flow itself.

\section{Observations}

The observations were made using NSFCAM on the IRTF in polarimetry mode. The basic technique is described in some detail in \citet{jon97}. Images are taken of the source and an off-source sky frame at four positions of a half waveplate that can be rotated in the beam. This allows us to form sky subtracted images at polarization position angles of 0, 45, 90 and 135$^\circ$. From these we computed Stokes images Q ($I_{0}-I_{90}$) and U ($I_{45}-I_{135}$) and the total intensity image $\frac{1}{2} (I_{0}+I_{45}+I_{90}+I_{135})$. Details of the flat-fielding and the image sequencing are given in \citet{jon97}.

Since NSFCAM is not a true imaging polarimeter, but rather a camera retrofitted with some polarization optics, we were limited in the precision of our polarimetry to about $\pm$0.3\% during our observations of DR21. This limit was due primarily to fluctuations in transmission and sky background. We know this to be the case because the minimum error can be improved by shortening the time between waveplate positions. Unfortunately to do so dramatically reduces our duty cycle and integration time on source. Since the nebulosity and stars in DR21 are faint, we kept the duty cycle at 80\%, which corresponds to about 12 seconds between waveplate positions.

Calibration of the efficiency and the position angle was done using observations of S1 in $\rho$ Oph with the same observing procedure as for DR21. S1 $\rho$ Oph was assumed to have a polarization of P = 1.95\% and $\theta$ = 28$^\circ$ at K (2.2$\mu $m). Instrumental polarization was checked by observing unpolarized standard stars from the UKIRT faint standards list and was found to be too small to be measured. These measurements also established our limiting precision for that night.

Seeing was about 1.1'' during our DR21 observations. Images were centered at 3 different locations in DR21. The Central image was centered on the bright stellar cluster at the heart of the DR21 complex. The approximate positions of the Eastern (E), Western (W) and Central images are given in Table 1. The sky background images for each of these positions were directly 2 arc minutes South. Since DR21 lies in a crowded region, the sky background images contained a significant number of field stars (but no nebulosity). These will show up in our images as black spots, since they produce negative values in the sky subtracted image. Although these stars could have been first 'removed' from the sky frames for cosmetic purposes, to do so would have caused confusion in our polarimetric data analysis. This is because it is unlikely we could have removed the effects of stars in the sky frames to a level of $\pm$0.3\% in polarimetric precision. By keeping the negative star images in the reduced data, we could more easily avoid measuring polarization in areas close to a subtracted field star from the sky image.

We measured the polarization of both the nebulosity associated with DR21 and a
number of stars in the images. Finding charts for the stars and locations of the measured nebulosity for the Eastern and Western lobes are shown in Figure 1. The measured polarizations of these objects are given in Tables 2 and 3. Results for the central image are given in Tables 4 and 5 where the positions are the offsets in arc seconds relative to the brightest source, IRS1, near the bottom center of the core cluster. 

The stars in all images were measured by using a set of concentric circular synthetic apertures with diameters of 2.4, 3.0 and 4.8''. The area between the outer two apertures was used to determine the nebular background which could then be subtracted from 'under' the image of the star in the inner aperture. This allows us to remove any polarized nebulosity from the stellar measurements, and is a major advantage of imaging polarimetry over aperture photopolarimetry. Results for 8 stars in the lobes are given in Table 2 and illustrated in Figure 2. Results for several stars and point-like objects in the central cluster are given in Table 4 and illustrated in Figure 4. 

Polarimetry of the nebulosity in the lobes was done using a 10'' square synthetic aperture. The results are given in Table 3 and shown in Figure 3. Note that we only have upper limits for locations in the Western lobe, which clearly has less polarized nebulosity than the Eastern lobe. Polarimetry of the nebulosity in the Central image was done using a 2.7'' square synthetic aperture. The results are given in Table 5 and illustrated in Figure 5. There were other locations in the images where we attempted to measure polarization but the nebulosity was too faint to yield any detections or any meaningful (5\%)upper limits. 

\section{Discussion}

\subsection{Interstellar Polarization}

ICBHT thoroughly discuss the issue of foreground interstellar polarization in the direction of DR21. Foreground polarization, if large, can contaminate the
polarization due to the dust associated only with DR21. To first order, this foreground polarization can be removed if it is known, but this is rarely the case. Despite a substantial effort, ICBHT were unable to satisfactorily determine the foreground polarization from their analysis of surrounding field star polarimetry from the literature. The surrounding field star polarimetry shows a rather chaotic distribution and is not very useful in determining the foreground polarization. 

The strength of the foreground polarization is probably low, however, based
on the magnitude of the polarization of nearby stars. ICBHT estimate a magnitude of 0.8\% at K attributable to foreground polarization. Unfortunately these optically visible field stars are all quite distant from DR21 on the sky, typically a degree or more. There are a few stars near the central cluster that \citet{dav96} identify as optically visible stars. These stars are probably the best candidates for foreground stars with only foreground polarization that are accessible in our images. Two of these are designated I4 and I5 by \citet{dav96} and are well enough separated from other sources for us to determine their polarization (Table 4). For both of these stars we have only an upper limit to the polarization of 0.9\% (3$\sigma$). These measurements support the conclusion by ICBHT that the foreground interstellar polarization is rather low.

The position angle of this weak (if any) foreground polarization is still undetermined. ICBHT investigated two possible cases for subtracting an interstellar polarization component, one parallel to the outflow and one perpendicular to the outflow. Neither case altered their conclusions. Since the position angle of the foreground polarization can not be determined by our observations nor by ICBHT, and the magnitude of the polarization is certainly less than 0.9\%, we will ignore it in this paper. We present other evidence later that further indicates foreground interstellar polarization is insignificant.

\subsection{Stars in the Central Cluster}

The polarization vectors for most stars in the central cluster (Figure 4) are in fair agreement with the bright central source, IRS1. Two stars, Nos. 8 and 9, have position angles more consistent with scattering than extinction and may be knots of reflection nebulosity. These sources are prominent in the broadband image taken by \citet{dav96}, but not their continuum subtracted H$_2$ image. This means that these two sources are not knots of H$_2$ emission but must be either stars or reflection nebulosity.

The position angle of the polarization of IRS1 (see Figure 2) is very nicely perpendicular to the 1.3mm polarization measured in a 19'' beam by \citet{gle99}. The millimeter polarization is due to emission of polarized light from aligned dust grains. If the same aligned grains emitting the mm flux are also causing the interstellar polarization in transmission at K, then the position angles would naturally be perpendicular to one another \citep{hil83}. Since the stars in the DR21 cluster are the source of energy for heating these grains, the mm emission is coming from dust local to DR21. Cold dust along the line of sight behind and in front of DR21 does not contribute to the mm polarization. Since our Near Infrared polarization of IRS1 has a position angle so close to perpendicular to the mm polarization, the extinction to DR21 must be dominated by grains local to the cluster, not in the foreground (unless the foreground interstellar polarization is at exactly the same position angle, of course).

This strongly suggests that at least for the central cluster, the grains causing the polarization we measure must be associated with DR21 itself. We can not make the same statement for the polarization in the lobes. The mm flux drops rapidly moving off from the central cluster and there are no published polarization measurements of thermally emitting dust in the lobes. This is unfortunate, because our probe of the magnetic field geometry in the lobes hinges on having the major fraction of the extinction to background stars associated with the lobes themselves.

Based on their narrow band imaging, \citet{dav96} suggest there is evidence for a second bipolar outflow from the IRS1 region with a North-South axis. We note that star No. 7 has a position angle that lies along the ridge of reflection nebulosity to the North of IRS1 and is suggestive of having a N-S component to the polarization in addition to the polarization seen towards IRS1. Whether or not this is evidence for a N-S flow with a magnetic field aligned in that direction is only speculation at this point. There are a number of stars further North than our observations could reach that will be the subject of future observations.

\subsection{Nebulosity in the Lobes}

The nebulosity in our bandpass consists primarily of scattered continuum light (presumably from the central cluster) and local H$_2$ emission. We were able to measure polarization of the nebulosity at 5 locations in the Eastern lobe (Figure 3). All five polarization vectors are perpendicular to the line of sight to the central cluster,  entirely consistent with simple reflection nebulosity if the illuminating source is IRS1 or some combination of stars in the central cluster. This implies that there is a relatively clear, extinction-free path between the central cluster and the dust in the Eastern lobe.

Perhaps surprisingly, we were unable to detect polarization in the Western lobe's nebulosity at the same level as seen in the Eastern lobe. Although our upper limits are only 4.2 and 6\% (3$\sigma$) in the Western lobe, polarizations as high as 21\% are present in the Eastern lobe. This suggests that the line of sight from the central cluster to the Western lobe may not be as clear and extinction free as in the East, and that most of the emission is from molecular hydrogen, relatively uncontaminated by reflection nebulosity. 

\subsection{Stars in the Lobes}

There are 8 stars in the lobes well isolated and bright enough for us to measure their polarization. All 8 stars have similar polarization position angles, basically aligned with the axis of the outflow (Figure 2). There is no indication of the presence of any magnetic field geometry along the line of sight to these stars other than aligned with the outflow. Note that the polarization is due to all of the dust along the line of sight, and our measurements represent a complicated weighted average of the effects of aligned grains along the line of sight \citep[cf.]{jkd}.

It is hard to see how this alignment of the polarization vectors with the outflow axis could be by chance. Even though we can not prove these stars are actually behind the outflow or that dust causing the polarization is primarily associated with the outflow itself, the data strongly suggest that it is so. There are three lines of circumstantial evidence that lead us to conclude that the polarization of the stars in the lobes is due to the local projected magnetic field geometry in the DR21 outflow. These arguments are:

\begin{enumerate}
    \item The polarization of stars in the central cluster is clearly from dust local to the DR21, not foreground dust (\S{3.2}).
    \item The nebulosity in the Eastern lobe shows no significant deviation from a simple scattering geometry even when the polarization strength is as low as 4.8\% (EN3), further indicating little contamination from foreground interstellar polarization.
    \item Interstellar polarization caused by dust not associated with DR21 (foreground and/or background) would have to be aligned with the outflow axis by chance. There is no indication in the surrounding field stars for such an alignment (ICBHT, \S{3.1}).
\end{enumerate}

ICBHT avoid some of this confusion by measuring the polarization of the H$_2$ emission in the outflow through a narrow band filter. This type of emission is local to the lobe, likely not scattered light from the central cluster, and therefore represents an intrinsically unpolarized source in the outflow itself. Consequently, there can be no contamination by dust from behind the lobes, only foreground interstellar polarization would be a concern (which we have argued is not a significant contributor). Our results for stars in the field are in excellent agreement with ICBHT in the Eastern lobe, where they also find polarization vectors aligned with the outflow axis. Clearly, the magnetic field geometry in the Eastern lobe is well aligned with the flow.

In the Western lobe, or results disagree with ICBHT, who find a polarization position angle perpendicular to the flow axis. Since we are observing stars, we can remove contamination from nebulosity in the star image (\S{2}). This makes our technique relatively less sensitive to polarized reflection nebulosity. ICBHT, however, were measuring nebular emission, and it is possible their results in the Western lobe were contaminated by reflection nebulosity in their beam. Scattering will, of course, produce a polarization perpendicular to the line-of-sight to the illuminating source, which is what they found. However, our inability to find highly polarized nebulosity through a broad band filter in the Western lobe would argue against this explanation. Either the ICBHT results are contaminated by reflection nebulosity in their beam, or there is a real difference between the magnetic field geometry in front of the H$_2$ knots they measured and in the dust along the lines of sight to the 3 stars we measured in the Western lobe. 

We find no evidence for a tightly wound spiral magnetic field in either lobe of the DR21 outflow. This does not necessarily mean that no such field geometry exists, it is just that the majority of volume associated with the extinction (dust) in the outflow does not contain such a geometry. It is possible that a spiral magnetic field is still responsible for the outflow and is present in the inner portions of the flow, but that there is relatively little dust optical depth associated with the volume occupied by the spiral field. Large amounts of dusty material entrained by the flow on its outside boundary could easily produce a cocoon around the flow that exhibits a magnetic field aligned with the outflow. More theoretical work needs to be done to determine if this could be the case.  

\section{Conclusions}

We have measured the polarization of stars and nebulosity in the field of the DR21 stellar cluster and in the Eastern and Western lobes of its massive bipolar outflow through a broadband K filter. Based on our observations we can reach the following conclusions: 

\begin{enumerate}
    \item If the stars are shining through the outflow from behind and there is minimal foreground and background interstellar polarization, then the magnetic field geometry sampled by the dust in the outflow is well aligned with the outflow axis. There is no evidence for a tightly wound spiral magnetic field nor significant fluctuations or turbulence in the field geometry. 
  \item The polarization vector of the bright central source IRS1 is nicely perpendicular to the polarized millimeter emission observed towards the same source. This strongly implies that the dust causing the K band polarization for IRS1 is local to DR21, not in the foreground.
  \item Through a broadband filter, the nebulosity in the Eastern lobe has significant polarization indicative of simple reflection nebulosity illuminated by the core cluster. Oddly, we were unable to measure polarization this strong in the nebulosity in the Western lobe. These results suggest that a clearer line of sight exists between the nebulosity in the Eastern lobe and the central illuminating cluster than is the case in the Western lobe.
   \item \citet{ito99} have made observations complimentary to ours. Our results agree with theirs in the Eastern lobe where we both find position angles aligned with the outflow axis. Our observations disagree with \citet{ito99} in the Western lobe where they find a position angle perpendicular to the outflow axis. Either the \cite{ito99} results are contaminated by reflection nebulosity in their beam, or there is a real difference between the magnetic field geometry in the H$_2$ knots they measured and the dust along the lines of sight to the stars we measured in the Western lobe.
\end{enumerate}

\section{acknowledgments}

We would like to thank the support staff of the IRTF for their assistance
helping make imaging polarimetry a viable option on the IRTF.

\newpage

\begin{deluxetable}{cccc}
\tablenum{1}
\tablewidth{0pt}
\tablecaption{Observing Log}
\tablehead{
\colhead{Image} & \colhead{RA (1950)} & \colhead{DEC (1950)} & \colhead{Filter}}
\startdata
Cluster Center & 20h 37m 14s & 42$^\circ$ 09' 00'' & K \\
Eastern Lobe & 20h 37m 22s & 42$^\circ$ 09' 15'' & K \\
Western Lobe & 20h 37m 08s & 42$^\circ$ 08' 30'' & K \\
\enddata
\end{deluxetable}

\newpage

\begin{deluxetable}{cccc}
\tablenum{2}
\tablewidth{0pt}
\tablecaption{Polarimetry of Stars in the Lobes}
\tablehead{
\colhead{Star} & \colhead{P (\%)} & \colhead{$\epsilon$P (\%)} & \colhead{$\theta$}($^\circ$)}
\startdata
E1 & 3.6 & 0.4 & 57 \\
E2 & 4.8 & 1.1 & 40 \\
E3 & 1.6 & 0.3 & 45 \\
E4 & 5.0 & 1.1 & 55 \\
E5 & 2.9 & 0.7 & 74 \\
W1 & 1.4 & 0.3 & 60 \\
W2 & 3.3 & 0.5 & 68 \\
W3 & 3.8 & 0.3 & 62 \\
\enddata
\end{deluxetable}

\newpage

\begin{deluxetable}{cccc}
\tablenum{3}
\tablewidth{0pt}
\tablecaption{Polarimetry of Nebulosity in the Lobes}
\tablehead{
\colhead{Star} & \colhead{P (\%)} & \colhead{$\epsilon$P (\%)} & \colhead{$\theta$}($^\circ$)}
\startdata
EN1 & 8.7 & 1.4 & 1 \\
EN2 & 10.1 & 1.5 & 174 \\
EN3 & 4.8 & 1.2 & 0 \\
EN4 & 6.4 & 1.2 & 170 \\
EN5 & 20.9 & 3.3 & 160 \\
WN1 & $\leq$6.0 & 2.0 & - \\
WN2 & $\leq$4.2 & 1.4 & - \\
WN3 & $\leq$6.0 & 2.0 & - \\
\enddata
\end{deluxetable}

\newpage

\begin{deluxetable}{ccccccc}
\tablenum{4}
\tablewidth{0pt}
\tablecaption{Polarimetry of Stars in the Central Cluster}
\tablehead{
\colhead{Star} & \colhead{RA ('')} & \colhead{DEC ('')} & \colhead{P (\%)} &
\colhead{$\epsilon$P (\%)} & \colhead{$\theta$($^\circ$)} & \colhead{Notes}}
\startdata
IRS1 & 0.0 & 0.0 & 6.8 & 0.3 & 107 & - \\
1 & -5.7 & 0.0 & 6.4 & 1.0 & 126 & - \\
2 & 0.3 & 14.3 & 7.5 & 0.6 & 125 & - \\
3 & -9.0 & 12.7 & 3.1 & 0.6 & 124 & - \\
4 & -21.0 & 11.0 & 3.7 & 0.6 & 90 & - \\
5 & -9.3 & -18.7 & 3.3 & 0.9 & 90 & - \\
6 & -8.3 & -31.7 & 2.5 & 0.7 & 119 & - \\
7 & -0.3 & 38.7 & 3.2 & 0.4 & 141 & - \\
8 & -2.0 & 7.0 & 6.3 & 1.1 & 85 & star? \\
9 & -3.0 & 4.3 & 6.4 & 1.3 & 75 & star? \\
I4 & -9.3 & -10.3 & $\leq$0.9 & 0.3 & - & Davis \& Smith (1996) \\
I5 & -12.3 & 3.0 & $\leq$0.9 & 0.3 & - & Davis \& Smith (1996) \\
\enddata
\end{deluxetable}

\newpage

\begin{deluxetable}{cccccc}
\tablenum{5}
\tablewidth{0pt}
\tablecaption{Polarimetry of Nebulosity in the Central Cluster}
\tablehead{
\colhead{Location} & \colhead{RA ('')} & \colhead{DEC ('')} & \colhead{P (\%)}
&
\colhead{$\epsilon$P (\%)} & \colhead{$\theta$($^\circ$)}}
\startdata
1 & 8.0 & 7.3 & 10.7 & 1.3 & 96 \\
2 & 5.3 & 10.3 & 10.2 & 1.3 & 101 \\
3 & 6.3 & 13.7 & 21.1 & 2.0 & 95 \\
4 & 6.3 & 16.7 & 17.9 & 2.0 & 110 \\
5 & 5.7 & 20.3 & 5.2 & 1.5 & 111 \\
6 & 3.7 & 23.0 & 4.6 & 1.7 & 126 \\
7 & 1.0 & 25.7 & 9.1 & 1.9 & 31 \\
8 & 41.0 & -9.0 & 5.8 & 1.2 & 91 \\
9 & 5.0 & 4.3 & 10.7 & 0.6 & 97 \\
10 & 7.0 & 2.3 & 12.2 & 0.6 & 100 \\
11 & 0.7 & 4.0 & 3.3 & 0.6 & 92 \\
12 & -2.7 & 2.0 & $\leq$ 2.1 & 0.7 & - \\
13 & -5.3 & 3.0 & $\leq$ 2.4 & 0.8 & - \\
\enddata
\end{deluxetable}

\newpage

\figcaption[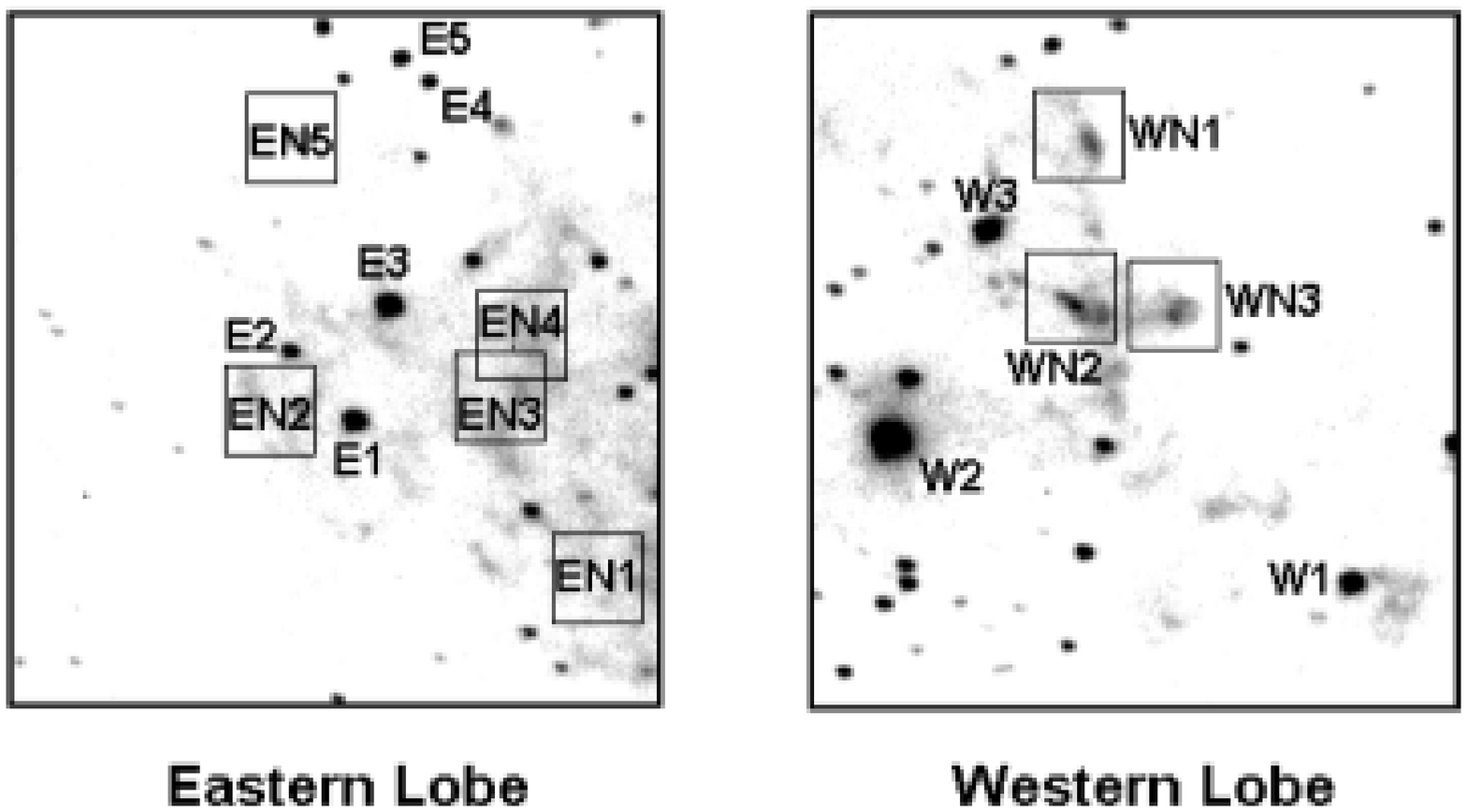]{Finding charts for the stars and nebulosity locations in the Eastern and Western lobes of the DR21 outflow (Tables 2 and 3). \label{fig1}}

\figcaption[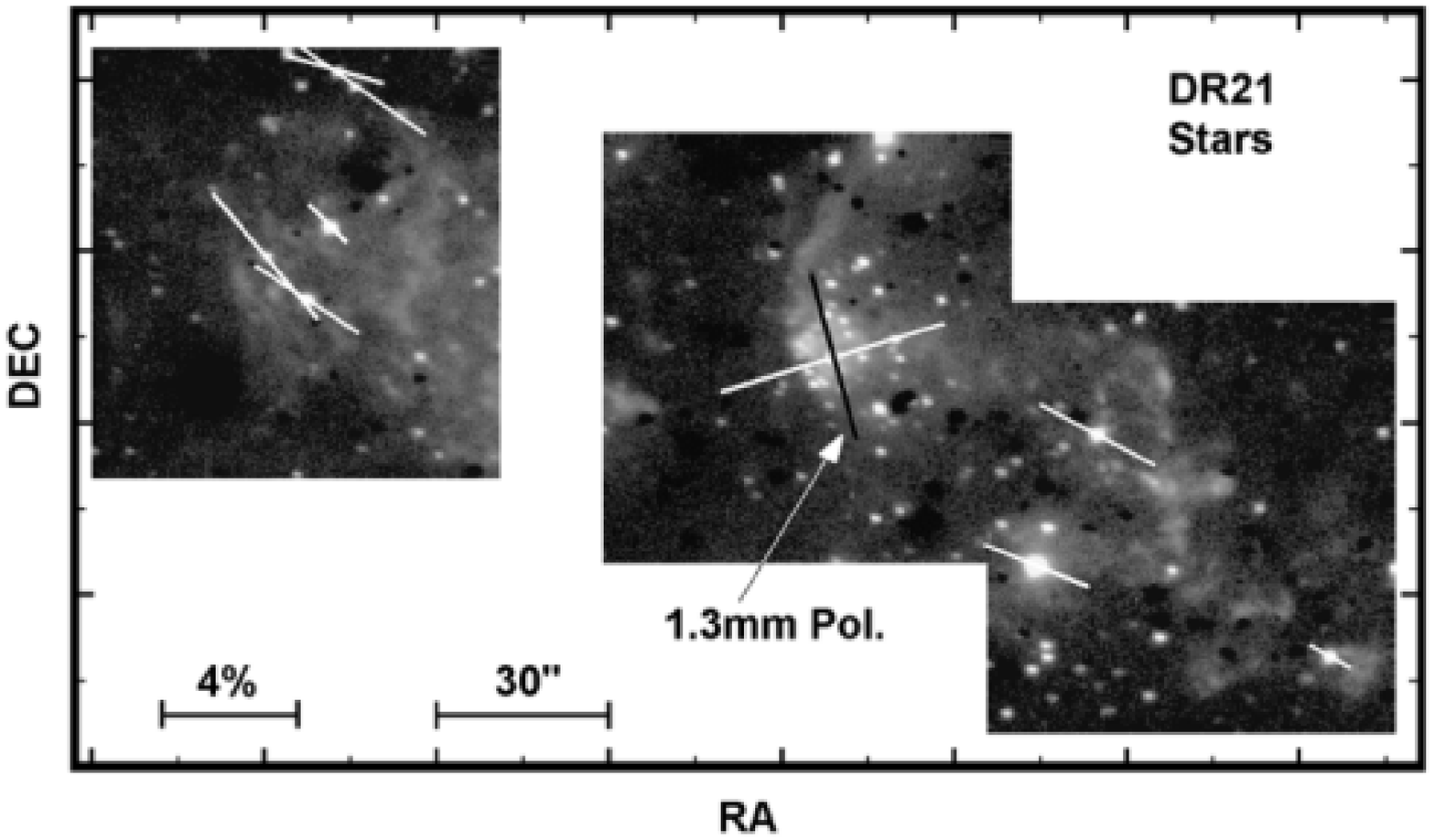]{Polarization vectors for stars shining through the bipolar outflow of DR21 (Table 2). Also shown is the polarization vector for the bright central source IRS1 and the millimeter polarimetry from \cite{gle99}. \label{fig2}}

\figcaption[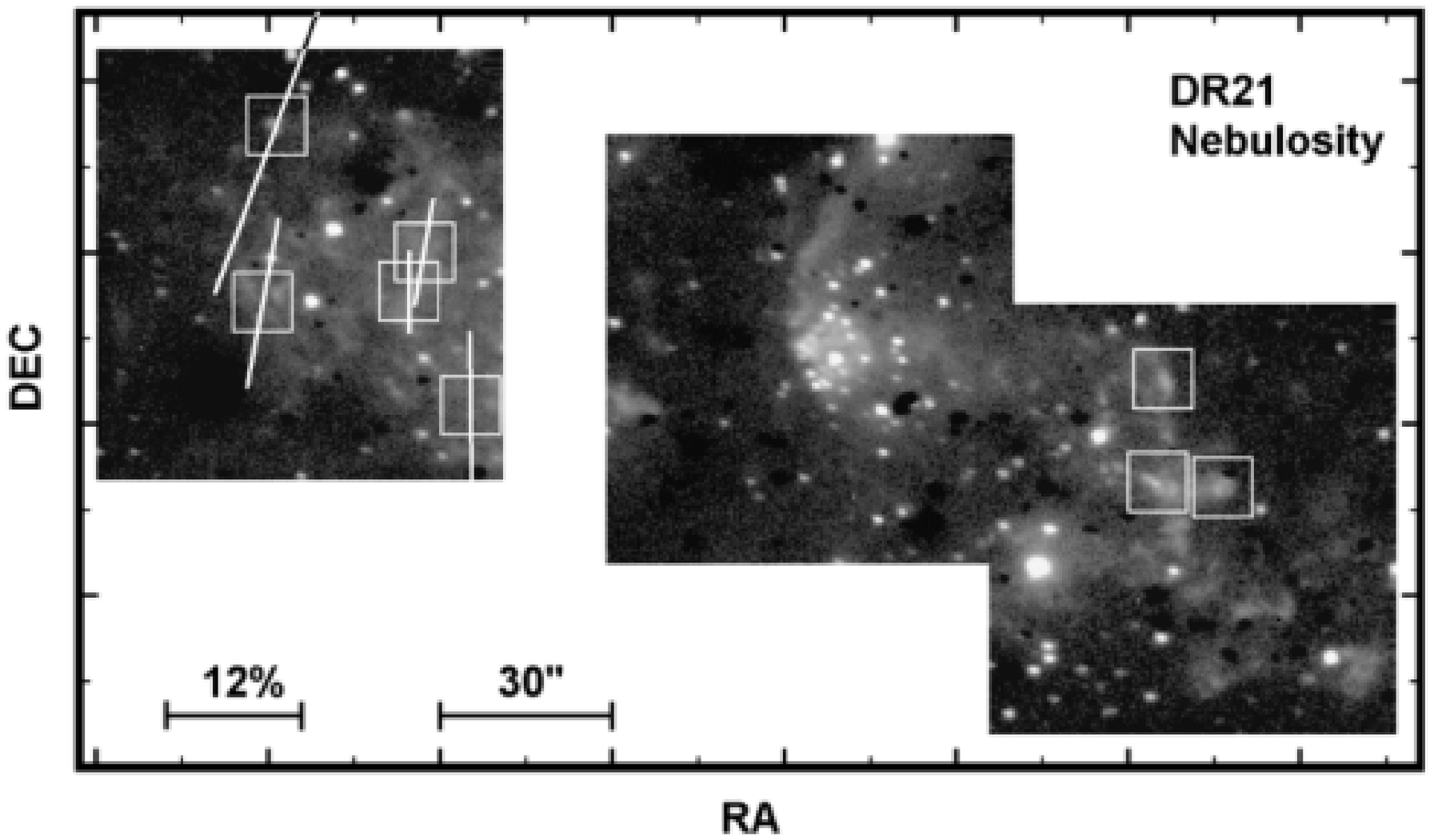]{Polarization vectors for nebulosity in the bipolar outflow of DR21 (Table 3). \label{fig3}}

\figcaption[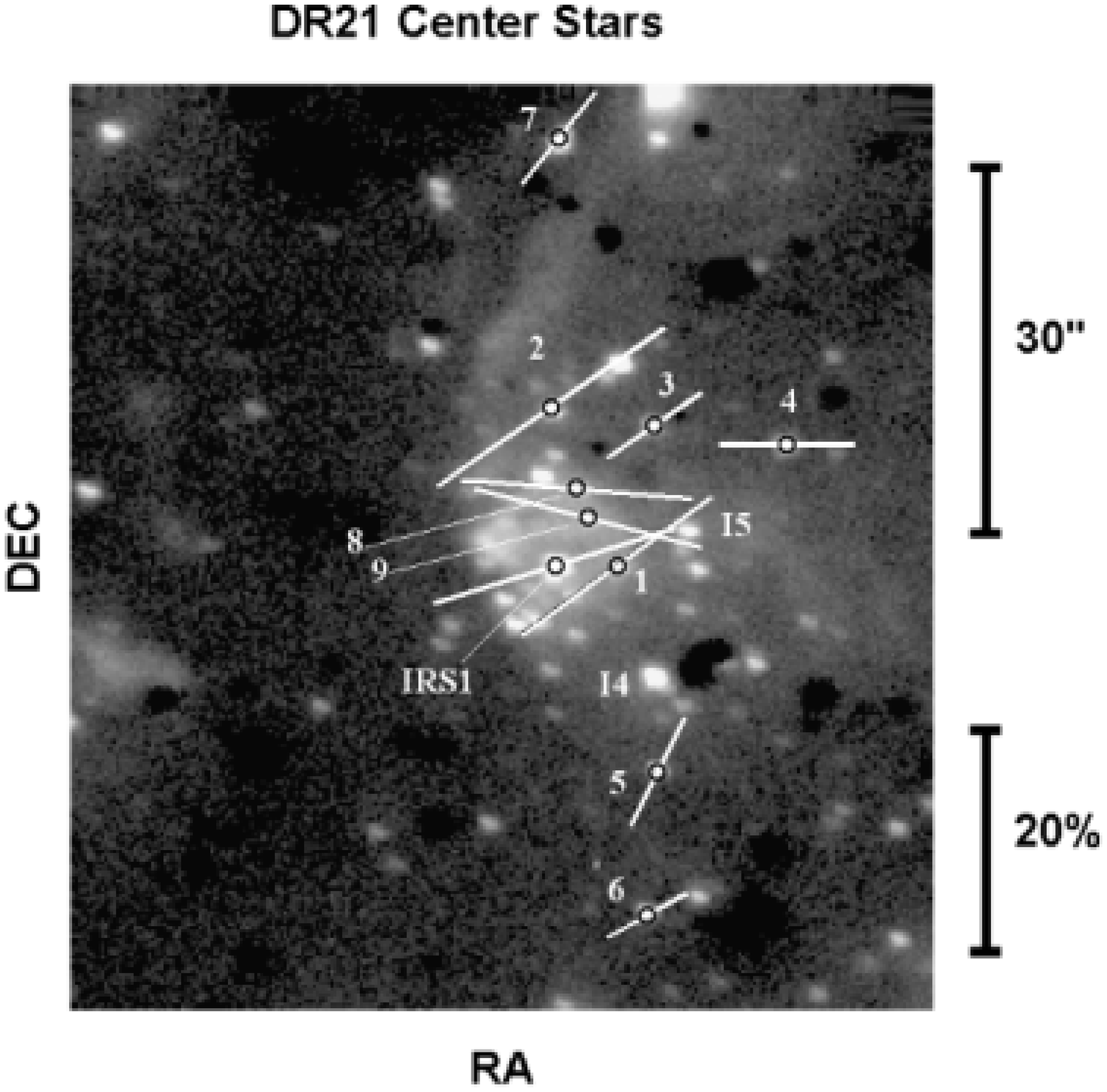]{Polarization vectors for stars in the central cluster (Table 4). \label{fig4}}

\figcaption[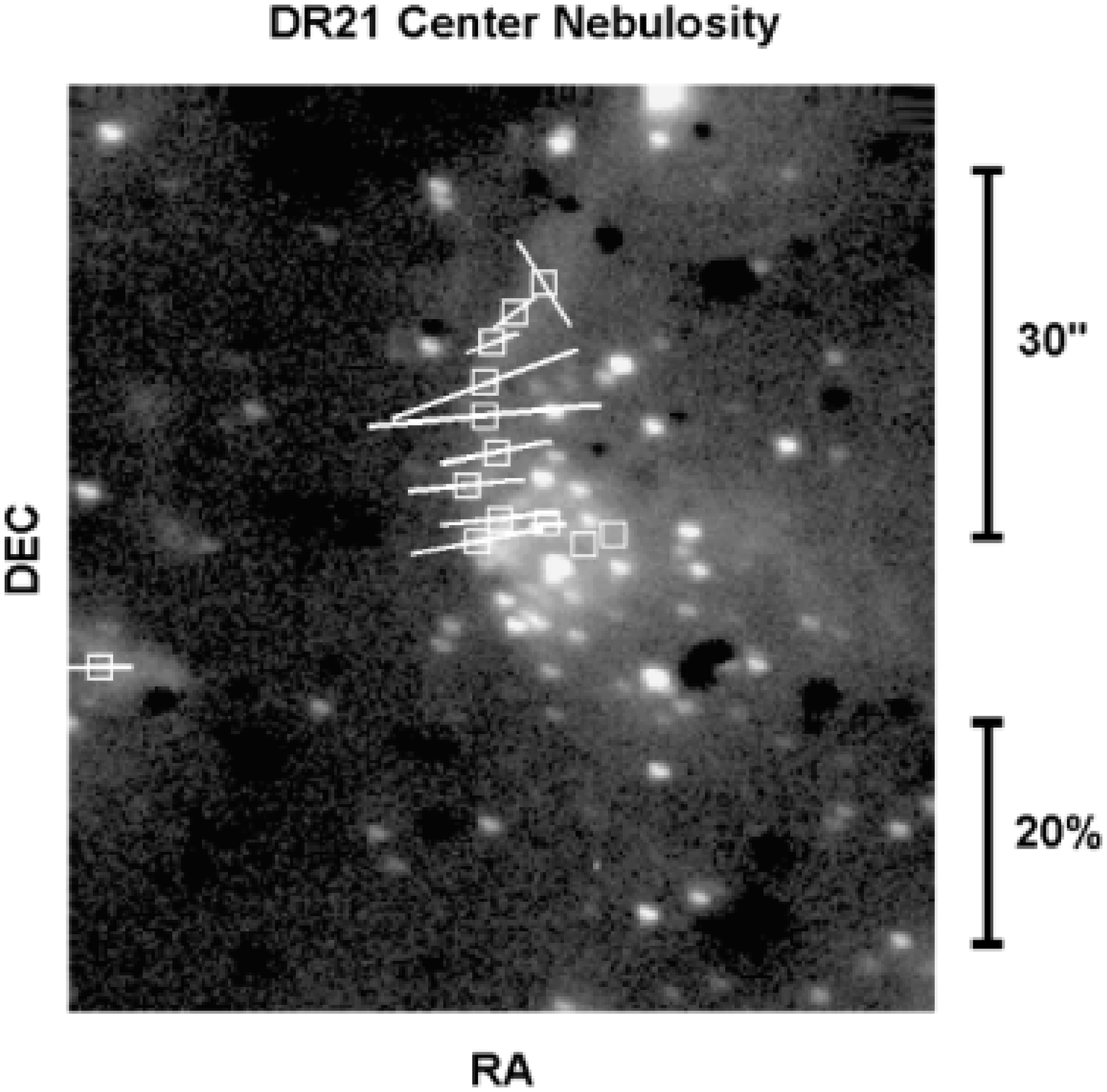]{Polarization vectors for nebulosity in the central cluster (Table 5). \label{fig5}}

\end{document}